\documentclass{aa} 

\usepackage{graphicx}
\usepackage{txfonts}
\usepackage{epstopdf}
\usepackage{natbib}

\usepackage{ulem}
\usepackage{xcolor}
\begin{document}

   \title{Shear instabilities in a fully compressible polytropic atmosphere}

   \author{V. Witzke \inst{\ref{inst1}}
          \and L. J. Silvers \inst{\ref{inst1}}
          \and B. Favier \inst{\ref{inst1}, \ref{inst2}}
          }

   \institute{Department of Mathematics, City University London,
              Northampton Square, London, EC1V 0HB, UK \\
              \email{Veronika.Witzke.1@city.ac.uk; Lara.Silvers.1@city.ac.uk}\label{inst1}
              \and
Aix-Marseille Universit\'{e}, CNRS, Ecole Centrale Marseille, IPHE UMR 7342, 49 rue F. Joliot-Curie, 13013 Marseille, France \email{Favier@irphe.univ-mrs.fr}\label{inst2}}



 
  \abstract
   {Shear flows have an important impact on the dynamics in an assortment of different astrophysical objects including accreditation discs and stellar interiors. Investigating shear flow instabilities in a polytropic atmosphere provides a fundamental understanding of the motion in stellar interiors where turbulent motions, mixing processes, as well as magnetic field generation takes place. Here, a linear stability analysis for a fully compressible fluid in a two-dimensional Cartesian geometry is carried out. Our study focuses on determining the critical Richardson number for different Mach numbers and the destabilising effects of high thermal diffusion. We find that there is a deviation of the predicted stability threshold for moderate Mach number flows along with a significant effect on the growth rate of the linear instability for small P\'eclet numbers. We show that in addition to a Kelvin-Helmholtz instability a Holmboe instability can appear and we discuss the implication of this in stellar interiors.
}

   \maketitle
%

\section{Introduction}
Understanding the complex dynamic interactions in the interior of stars, such as the Sun, is crucially important if we are to develop a physical model of these objects in their entirety. To begin to obtain a comprehensive knowledge of the motions in stars it is convenient to focus initially on the Sun, which we have the most detailed observational evidence for. Helioseismology has shown that at the base of the solar convection zone there is a thin region of radial shear called the tachocline \citep{Kosovichev1997, Tobias_book}. This region is believed to play a crucial role in the solar dynamo \cite[see][and references therein]{2008RSPTA.366.4453S}. However, in spite of the evidence of the existence of the tachocline and its importance, there is still a considerable amount of work to be undertaken to understand this region using mathematical modelling techniques.\\
Velocity measurements suggest that the tachocline region is hydrodynamically stable against vertical shear flow \citep{CBO9780511536243A016}. However, helioseismology is restricted  to large-scale time averaged measurements \citep{CBO9780511536243A013} and so turbulent motions can be still present on small length and time scales. Thus it is very plausible for the tachocline to appear to be stable, using current helioseismology techniques, but actually to be hydrodynamically or magnetohydrodynamically unstable. Though it is widely assumed that the tachocline is stable \cite[see][]{Tobias_book}, \citet{2000A&A...364..876S} have shown that shear turbulence can appear in a narrow part of the tachocline. An unstable tachocline would be significantly different in its dynamical interactions from a stable region and so, if we are to understand the role of this region, for example in the solar dynamo we must first understand unstable shear flows in a polytropic atmosphere.\\
Shear flows occur in a wide variety of natural settings as for example in oceanic flows, planetary atmospheres, stars and galactic discs. Therefore, there have been a number of previous investigations that examine shear flows in different contexts that can help inform our approach to the examination of shear flows in stars.\\
Previous studies of shear flows have shown that such flows can undergo what is known as the Kelvin-Helmholtz (KH) instability, which develops due to conversion of the available kinetic energy of the shear flow into kinetic energy of the disturbances  \cite[see][chap.~6]{Drazin_Reid_book}. In addition to the KH instability, other instabilities such as baroclinic instability \cite[see][]{baroclinic_01}, or the Holmboe instability \citep{Holmboe_1962}, can appear when flows are either rotating or stratified. For our study the latter one is of greater interest because it is known that, while the KH instability is suppressed by stratification, the more slowly growing Holmboe modes become dominant with increasing stratification \citep{Smyth_peltier_holmboe}.\\
To study any kind of instability it is convenient to start with investigating the stability threshold of the system. 
For the extensively studied KH instability, the necessary criterion for stability requires the Richardson number to be greater than $1/4$ everywhere in the domain \citep{Miles1961}. This criterion was derived for simplifying assumptions, where the fluid is incompressible, inviscid and non-diffusive.
However, dropping these simplifications may alter the stability criterion such that in a system where thermal diffusion becomes important, and acts on a smaller time scale than buoyancy, the stability criterion requires a significant modification. \citet{Dudis_1974} and \citet{Zahn_1974} have shown that in such systems the product of the Richardson number with the P\'eclet number is the quantity that indicates stability. The effect of thermal diffusion on shear instabilities was only studied in the Boussinesq approximation by \citet{Jones1977}, \citet{Dudis_1974} and more recently by  \citet{1999AA...349.1027L}, such that it is not directly applicable for stellar interiors where large pressure gradients have to be considered. In a general fully compressible model there is the potential for the stability criterion to be altered as the Mach number is varied. In most stellar regions the Mach number is assumed to be small but it can still be potentially significant. One example are coronal mass ejections where shear flow instabilities were observed recently by \citet{2041-8205-734-1-L11}.\\
\citet{dissertationMiczek} considers a fully compressible fluid in an adiabatic atmosphere, but the effect of varying all, especially thermal, transport coefficients was not studied. Therefore, this study does not capture all relevant effects present in stellar interiors. 
Although, considerable work has been undertaken to examine shear flows in a variety of different contexts, no work to examine shear flows in a polytropic atmosphere has been carried out and thus will be what we investigate here.\\
In this paper we conduct a linear stability analysis to examine both the effect of high thermal diffusion and the effect of compressibility on the onset of shear flow instabilities in a stably stratified polytropic atmosphere. While the main focus is on KH instabilities the appearance and consequences of a Holmboe like instability is investigated. The governing equations are given in Sect. \ref{M_Model} along with the numerical method used. Our results are presented in Sect. \ref{R_Result} followed by a discussion in Sect. \ref{C_Conclusion}.
\section{Model}
\label{M_Model}
\subsection{Governing equations, boundary conditions and background state.}
\label{sec:Governing_equ}
We consider a compressible fluid in a Cartesian domain bounded at $z=0$ and $z=1$ and periodic in x and y directions. 
The fluid is assumed to be an ideal gas with constant dynamic viscosity, $\mu$, constant thermal conductivity, $\kappa$, constant heat capacities $c_p$ at constant pressure and $c_v$ at constant volume.
The equations we consider, in non-dimensional form, are
\begin{eqnarray}
\label{eq:NSEquation01}
\frac{\partial \rho}{\partial t} & = & - \mathbf{\nabla}\mathbf{\cdot}\left(\rho \mathbf{u} \right) \, \\
\frac{\partial(\rho \mathbf{u})}{\partial t} & = & \sigma C_k \left( \nabla^2\mathbf{u}\, +\,\frac{1}{3}\mathbf{\nabla}(\mathbf{\nabla}\mathbf{\cdot}\mathbf{u})\right) -\mathbf{\nabla} \mathbf{\cdot} \left(\rho \mathbf{u u} \right)\, \nonumber \\
 &   & -\,\mathbf{\nabla}p\, + \, \theta(m+1) \rho\, \hat{\mathbf{z}} \, \\
\label{eq:NSEquation02}
\frac{\partial T}{\partial t}  & = & \frac{C_k \sigma (\gamma -1)}{2\rho}|\mathbf{\tau}|^{2}\,+\,\frac{\gamma C_k}{\rho} \nabla^2 T \nonumber \\
 &  & - \mathbf{\nabla} \mathbf{\cdot}\left(T \, \mathbf{u}\right)\,-\,(\gamma -2)T \mathbf{\nabla} \mathbf{\cdot} \, \mathbf{u}
\label{eq:NSEquations}
\end{eqnarray}
where $\rho$ is the density, $\mathbf{u}$ the velocity field, $T$ the temperature, $\theta$ denotes the temperature gradient, and $p$ is the pressure. The dimensionless Prandtl number, $\sigma=\mu c_p/\kappa $, is the ratio of viscosity to thermal conductivity, the thermal dissipation parameter is defined as $C_k= \kappa \tilde{t}/(\rho_0 c_p d^2)$ and $\gamma= c_p/c_v$ denotes the adiabatic index. The strain rate tensor has the form 
 \begin{equation*}
 \tau_{ij}=\frac{\partial u_{j}}{\partial x_i} + \frac{\partial u_i}{\partial x_j} - \delta_{ij} \frac{2}{3}  \frac{\partial u_{k}}{\partial x_{k}} .
\end{equation*} 
In the dimensionless equations above, all lengths have been scaled with the domain's depth $d$. Recasting the temperature and density in units of $T_{t}$ and $\rho_{t}$, the temperature and density at the top of the layer, and taking the sound-crossing time, which is given as $\tilde{t} = d/[ (c_p-c_v)T_{t} ]^{1/2}$, as the fundamental time it follows that the pressure $p$ is given in units of $p_{t} = (c_p- c_v) \rho_{t} T_{t} $ and the velocity field is given in units of acoustic wave velocity.\\
For the background state we assume a polytropic relation between pressure and density such that the pressure is a function of density only i.e.
\begin{equation}
p(\rho)\,\propto \,\rho^{\left(1+\frac{1}{m}\right)}
\end{equation} 
where $m$ is the polytropic index. Note, this relation is not valid for the perturbed quantities derived in the next section. Due to the Schwarzschild criterion the fluid is stable against convection if the inequality $m > 1/(\gamma -1)$ holds, which is the case for a polytropic index of $m > 1.5$. Only stable stratified atmospheres will be considered throughout the paper.\\
Boundary conditions, at the top and bottom of the of the domain, are impermeable and stress-free velocity i.e. 
\begin{equation}
{u}_z = \frac{\partial {u_x}}{\partial z} = \frac{\partial {u_y}}{\partial z} = 0 \quad \text{at} \quad  z= 0 \quad  \text{and} \quad z = 1
\label{eq:boundary01}
\end{equation}
and fixed temperature at the top and bottom:  
\begin{equation}
T=1 \quad \text{at} \quad z=0 \quad \text{and} \quad T=1+\theta \quad \text{at} \quad  z=1. \quad
\label{eq:boundary02}
\end{equation} 
To include thermal effects it is necessary to choose the background temperature in such a way, that it is a stationary solution of the heat equation or remains quasi-stationary on time scales larger than the thermal diffusion time scale. This results in a temperature and density profile of the form:
\begin{equation}
T(z)\,=\, T_{t}\left(1 \,+\, \theta z \right)
\label{eq:initialTemp}
\end{equation}
\begin{equation}
\rho (z) = \rho_{t} \left( 1 + \theta z \right)^{m}
\label{eq:initialdens}
\end{equation}
where $\theta$ is the dimensionless temperature difference between the upper and lower boundaries of the domain. These equations form only an equilibrium state if the fluid is at rest or viscous heating is negligible. The background velocity profile takes the form
\begin{equation}
 u(z)= U_0 \tanh\left(\frac{z-0.5}{L_u}\right)\,
 \label{eq:shearprofile}
 \end{equation}
with a shear amplitude $U_0$ and a scaling factor $1/L_u$ that controls the width of the shear profile. 
\begin{figure}
   \centering
   \includegraphics[width=0.95\hsize]{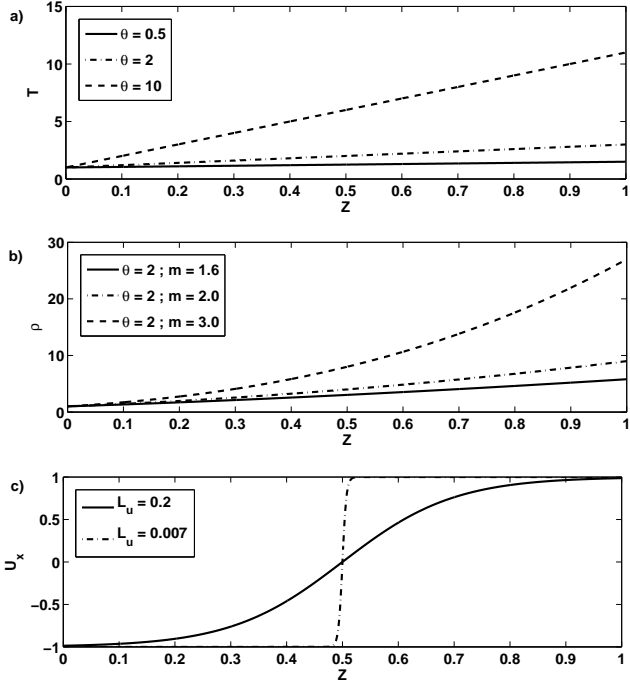}
      \caption{Plots of the typical background density, temperature and shear flow profiles. a) Temperature profiles for different $\theta$, that were used. b) Density profiles for $\theta=2$ but three different polytropic indices $m$. For all indices $m$ the atmosphere is stably stratified.  c) Shear flow profiles with the smallest and largest characteristic length $L_u$ used are shown. }
         \label{fig:back_profile}
   \end{figure}
The boundary conditions introduced in equation (\ref{eq:boundary01}) and (\ref{eq:boundary02}) restrict the shear profile to values of $L_u$ which will result in a small enough value of the z-derivative at the boundaries. For the static state the temperature and density profiles are taken as in equation (\ref{eq:initialTemp}) and equation (\ref{eq:initialdens}), respectively. Then, the equilibrium state is characterised by $\mathbf{u}_0(z) =(u(z),0,0)^T $, $T_0(z)$, $p_0(z)$ and $\rho_0(z)$. Selected background profiles for temperature, density and velocity are shown in Fig. \ref{fig:back_profile}.
\subsection{Formulation of the Eigenvalue problem}
In a diffusive model there are a number of different time scales including the time scale associated with the shear dynamics $t_S=L_u/(U_0)$, the time scale for buoyancy $t_B = 1/N(z)$, where $N(z)^2$ is the Brunt-V{\"a}is{\"a}l{\"a} frequency, and the time scale for thermal diffusion $t_{k} = L_u^2/C_k$.  In this paper, we focus on the regime where the viscous time scale, $ t_{\mu} = L_u^2/ \mu$, is much greater than any other time scales. This allows us to neglect viscous heating, which corresponds to the first term on the right hand side in equation (\ref{eq:NSEquations}). In addition, the shear flow given by equation (\ref{eq:shearprofile}) is in equilibrium only if $t_{\mu}$ is much greater than the instability time scale, which we verify \textit{a posteriori}.  When $t_{\mu}$ becomes comparable with other relevant time scales, the background shear flow is not in equilibrium and our analysis would be inappropriate for this case.\\
We perturb each quantity that appears in equations (\ref{eq:NSEquation01}) - (\ref{eq:NSEquations}) such that $f=f_0+\delta f$ and 
\begin{equation}
\delta f (x,y,z,t)  = \tilde{f}(z) \exp{\left(ikx + ily +\zeta t \right)},
\end{equation}
where $k \in \mathbb{R}$ and $l \in \mathbb{R}$ are the horizontal wave numbers, and $\zeta = \zeta_r + i \zeta_i \in \mathbb{C}$, where $\zeta_r$ gives the growth rate of the linear instability.\\
Note, the equations for the perturbed quantities that we obtain do not inherit the same symmetry properties as the well known Taylor-Goldstein equation (e.g., \cite{Miles1961}), where taking the complex conjugate of the eigenfunction and eigenvalue leads to the same equation.
This symmetry is broken in our set of equations, because there are still terms linear in $k$ and $l$. 
Therefore, for our eigenvalue problem there do not necessarily exist two complex conjugated solutions where one is decaying and one is a growing solution.\\
For our initial set of equations a Squire transformation  exists to transform the three-dimensional problem to a corresponding two-dimensional one.  The transformation can be written as:
\begin{eqnarray}
\tilde{k}^2 & = & k^2 +l^2 \qquad \tilde{k} \delta \tilde{u} \, = \,  k \delta u + l \delta v \qquad \delta \tilde{w}  =  \delta w \nonumber \\
\tilde{U_0} &  = & U_0 \qquad \tilde{c}  =  c \qquad \delta \tilde{\rho} = \frac{\tilde{k}}{k} \frac{\theta \left(m+1\right)}{\tilde{\theta} \left(\tilde{m} +1 \right)} \delta \rho  \nonumber\\ 
\delta \tilde{T} & = & \frac{\tilde{k}}{k} \delta T \qquad \tilde{T_0} = \frac{\tilde{k}^2}{k^2} T_0 \qquad \nonumber \\
{\tilde{\rho}_0} & = & \frac{{k}}{\tilde{k}} \frac{\tilde{C}_k}{C_k} {\rho_0}  \qquad \delta\tilde{ \rho} = \frac{k}{\tilde{k}} \frac{{\tilde{\rho}_0} }{\rho_0} \delta \rho
\end{eqnarray}
Having also checked numerically that indeed for a certain wave number $k$ the growth rate $\zeta_r$ decreases with increasing $l$, we can set $l=0$ without loss of generality for our following computations. Denoting $\delta \mathbf{u} = (u,v,w)$, we obtain this linearised coupled set of equations   
\begin{eqnarray}
\zeta \delta \rho &=& -iku\rho_0 -\frac{\partial}{\partial z}(w\rho_0) -ik \delta \rho U_0\\
\label{eq:EVProbelmsystem01}
\zeta \rho_0 u &=& -ik\left(\rho_0 \delta T +T_0 \delta \rho\right) -\rho_0 w \frac{\partial U_0}{\partial z} -ik\rho_0 u U_0 \nonumber \\
           & & - C_k \sigma \left(\frac{4}{3}k^2 u -\frac{\partial^2 u}{\partial z^2} -\frac{1}{3}ik\frac{\partial w}{\partial z} \right)\\
\label{eq:EVProblemsystem02}           
\zeta \rho_0 w &=& -\frac{\partial}{\partial z}\left(\rho_0 \delta T +T_0 \delta \rho  \right)  -ikU_0 \rho_0 w + \theta \left(m +1 \right)\delta \rho \nonumber \\
           & & - C_k \sigma \left(k^2 w -\frac{4}{3} \frac{\partial^2 w}{\partial z^2} -\frac{1}{3} \frac{\partial}{\partial z} \left(iku \right) \right) \\
\label{eq:EVProblemsystem03}           
\zeta \delta T &=& \frac{C_k \gamma}{\rho_0}\left(  \frac{\partial^2 \delta T }{\partial z^2} -  k^2 \delta T \right)  -ikU_0 \delta T \nonumber \\ 
           & & - \left( \gamma -1 \right) T_0 \left( iku + \frac{\partial w}{\partial z} \right) - w \frac{\partial T_0}{\partial z},\,
\label{eq:EVProbelmsystem}
\end{eqnarray}
where a similar set of equations was derived for a problem including magnetic fields by \citet{2004ApJ...603..785T}. Our system is characterised by six parameters $m$, $\theta$, $\sigma$, $C_k$, $ U_0$ and $L_u$. 
Equations (\ref{eq:EVProbelmsystem01}) - (\ref{eq:EVProbelmsystem}) are numerically solved on a one dimensional grid in $z$-direction that is discretised uniformly, this method is adapted from the method used by \citet{2012MNRAS.426.3349F}. Recasting the set of differential equations into the form
\begin{equation}
\zeta  \mathbf{f} = A  \mathbf{f},
\end{equation}
where the matrix $A$ contains the finite difference coefficients applied to the discretised eigenfunctions $ \mathbf{f} =(\delta \rho, u, v, w, \delta T)^T$, reduces the problem to a matrix equation. To find the eigenvalues and vectors the Schur factorisation is used \citep{anderson1999lapack}. For the computation of the relevant coefficients in $A$, a central fourth-order finite differences scheme was used. We search for the eigenvector solutions with the greatest real part of the eigenvalue $\zeta$ and where the vertical velocity eigenvector, $w$, vanishes at the boundaries.
Ultimately, we aim at undertaking non-linear simulations with a pseudo-spectral code and viscosity will be mandatory in that case. Therefore, most of the computations will consider a viscous fluid. 
%
\section{Results}
\label{R_Result}
In this section we focus on a number of key areas of interest. First, we present the change of the stability threshold while the Mach number is varied, which correspond to a continuous transition between an incompressible and a compressible fluid. The effect of compressibility is separately investigated in a weakly thermally stratified and a strongly thermally stratified atmosphere Sect. \ref{subsec:low_temperature_gradient}. Later in Sect. \ref{subsec:Peclet_number_var}, the growth rates of the linear shear instability together with the critical Ri for different P\'eclet numbers are compared and the effect on the stability against buoyancy is discussed. In Sect. \ref{sec:varying_m} the effect of different polytropic indices on the instability is addressed and the  possibility of a Holmboe like instability is investigated in Sect. \ref{subsec:Holmboe}.
\subsection{The effect of varying the Mach number on the instability threshold}
\label{subsec:low_temperature_gradient} 
\begin{figure}
   \centering
   \includegraphics[width=\hsize]{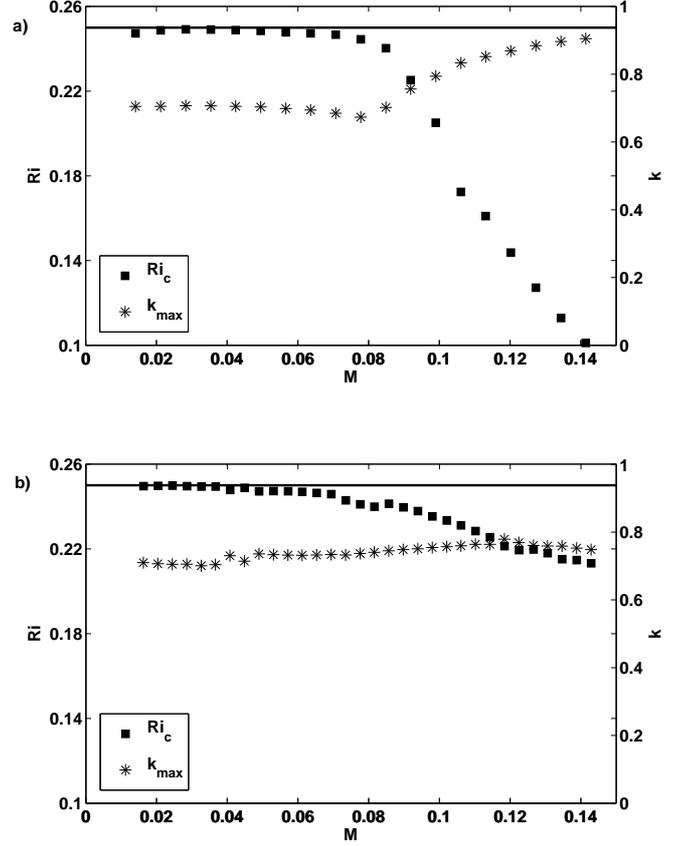}
      \caption{
     In both plots the critical Richardson number is found in a viscous fluid for different Mach numbers, M.  The corresponding wave numbers $k_{max}$ of the most unstable mode at the onset of instability are plotted, the wave number is normalised by the inverse of the characteristic length $1/L_u$. The horizontal line in both plots correspond to $Ri=0.25$. In a) the atmosphere is weakly stratified with $\theta = 2$ whereas in b) $\theta = 10$ which corresponds to a strongly stratified atmosphere. }
         \label{fig:crit_Ri_with_mach}
   \end{figure}

\begin{figure}
\centering
\includegraphics[width=\hsize]{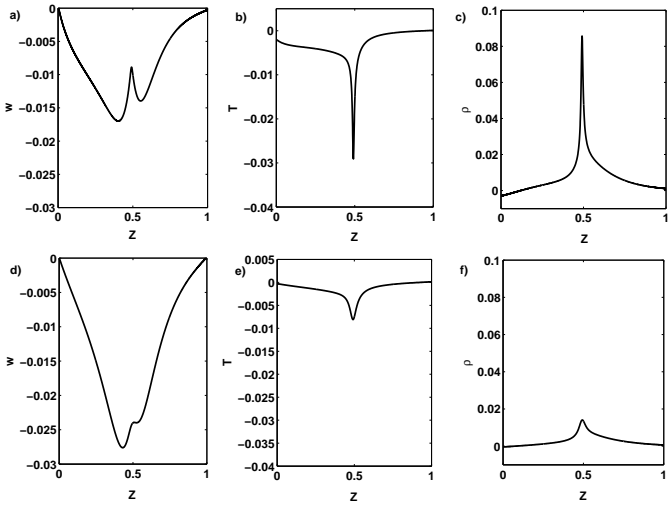}   
    \caption{Eigenfunctions for a fixed $M= 0.114$ and a fixed characteristic length scale $L_u = 0.09$ are found for two different $\theta$, where $\theta = 2$ for the plots d, e, f and $\theta =10 $ for top plots a, b, c. In both cases the eigenfunctions for the most unstable mode are shown.}
         \label{FigVibStab}
   \end{figure}

As the Richardson criterion is based on simple energetic arguments and does not take compressibility into account, clarification is needed to determine whether compressibility affects the stability of a shear flow. Therefore, in this section we focus on the stability threshold for different Mach numbers in a viscous (we consider $\sigma=1.0$ and $C_k = 10^{-6}$) and stably stratified fluid with $m =1.6$. Although stellar interiors have typically low Mach numbers, especially at the base of the convection zone, generally moderate to high Mach numbers can appear at the surface and in other astrophysical objects. Thus, investigating the consequences of moderate Mach numbers on a shear flow is of general interest.
In the following we refer to the Mach number, M(z), as 
\begin{equation}
M(z)= \frac{U_0}{\sqrt{1+\theta z}}.
\label{eq:Machnumber_domain}
\end{equation}
This is the consequence of our previous definition, where velocity is given in units of the sound speed that is computed at the top of our domain.
As the inflexion point of our shear flow is at $z=0.5$, and the sound speed varies with temperature, it is necessary to compute the actual Mach number, M, at $z=0.5$.\\
According to \citet{schochet1994fast} and \citet{Guillard2004655}  the solutions of the compressible Euler equations reduce to the solutions of the incompressible Euler equations in the low Mach number limit.  Thus, varying the Mach number allows to investigate the validity of the Richardson criterion for low to moderate Mach numbers ($0.02 < M < 0.15$). \\
We make use of the general definition of the Brunt-V{\"a}is{\"a}l{\"a} frequency given by
\begin{equation}
N^2(z) =  \frac{g}{\tilde{T}} \frac{\partial \tilde{T}}{\partial z}, \,
\end{equation}
where $\tilde{T} = (P_{t}/P)^{1-1/\gamma} $ is the potential temperature,  to define the local Richardson number as
\begin{eqnarray}
Ri_{min} & = & \min_{0 \le z \le 1} \left(N(z)^2 \left/ \left( \frac{\partial u(z)}{\partial z} \right)^2 \right.  \right) \nonumber \\
 & = & \min_{0 \le z \le 1}  \left( \frac{\theta^2 (m +1) \left(\frac{m+1}{\gamma} - m\right)}{\left(1+\theta z \right) \left(\frac{U_0 - u(z) }{L_u}  \right)^2 } \right),
\label{eq:Richardsonnumber_def}
\end{eqnarray} 
where the derivative of the background velocity profile with respect to $z$ corresponds to a local turnover rate of the shear.\\
To find the critical Richardson number, $Ri_c$, we solve the eigenvalue problem for a small $Ri$, while varying the wave number, $k$, between $0$ and $1/L_u$ to find the most unstable mode $k_{max}$.
For large, but finite, Reynolds number the system is assumed to be stable if the growth rate, $\zeta_{r}$, is zero for all wave numbers or the time scale for the instability, $t_{\zeta}= 1/\zeta_{r}$, compared to the viscous time scale, $t_{\mu}$, is of the same order. \\
A detailed survey for $\theta = 2$, a weakly stratified atmosphere, and for $\theta = 10$, a strongly stratified atmosphere, reveals that the critical Richardson number decreases for Mach numbers greater  than $0.08$, which can be seen in Fig. \ref{fig:crit_Ri_with_mach}. For the majority of incompressible, and weakly compressible Mach numbers the critical Richardson number does not significantly deviate from the well known $1/4$ threshold for stability. In the case of a weakly stratified atmosphere the critical Richardson number decreases rapidly below $0.1$ for $M \approx 0.14$. In a strongly stratified fluid we find qualitatively the same behaviour as for the weakly stratified case but the critical Richardson number does not drop below $0.2$ for $M \approx 0.14$. The shift of the stability threshold, in both a strong and a weakly stratified fluid, towards smaller Richardson numbers for moderate Mach numbers indicates a stabilising effect of compressibility on the KH instability. Though not only focusing on KH like instabilities, previous investigations of compressible shear flows with uniform temperature, and without gravity, for high Mach numbers revealed similar results. For example \citet{FLM:386771} and \citet{FLM:387688} showed that for $M \rightarrow \sqrt{2}$ the system becomes stable. In this system two types of unstable modes are present for Mach numbers greater than $0.94$, which are stationary modes and travelling modes.\\ 
In the case of KH like instabilities one explanation for the stabilising effect for Mach numbers greater than $0.08$ is as follows. The Richardson criterion uses simple energetic arguments, where two neighbouring fluid parcels are exchanged. 
The density of these parcels remains constant for an incompressible fluid such that only the change in velocity (at different heights) changes the kinetic energy,  $\Delta E_{kin}$, and changes in the potential energy, $\Delta E_{pot}$, are solely due to the changes in position of the fluid parcel. While for a compressible fluid the density will decrease to a certain amount when a fluid parcel is moved up adiabatically, such that a part of the kinetic energy is converted by the process of expansion. Therefore, to reach the instability threshold more kinetic energy is needed which requires a greater velocity gradient. \\
Different temperature gradients have a non-trivial effect on the effective Mach number throughout our domain, where $M$ changes according to equation (\ref{eq:Machnumber_domain}). In the limit of weak thermal stratification $M$ remains almost constant whereas in a strongly stratified atmosphere it changes significantly and generates an asymmetry between the regions above and below the shear flow.
The observed asymmetry changes the eigenfunctions found for a fixed $M (z = 0.5)$  and $L_u$, which are shown in Fig. \ref{FigVibStab}. Note that because $M (z = 0.5)$ and $L_u$ are fixed, increasing $\theta$ is equivalent to increasing the Richardson number (see equation (\ref{eq:Richardsonnumber_def})). The asymmetry with respect to the mid-plane is clearly visible in the temperature and vertical velocity eigenfunctions for the strongly-stratified case $\theta = 10$ (see Fig. \ref{FigVibStab} a and b) whereas the same eigenfunctions are much more symmetric for $\theta = 2$ (see Fig. \ref{FigVibStab} d and e). 
This asymmetry has impact on the observed deviation between the change of the critical Richardson number for $\theta=2$ and $\theta=10$. In fact, in a stratified atmosphere exchanged fluid parcels, where one is shifted downwards form the middle plane $z=0.5$ and the other upwards, move at the same speed in opposite directions, but have different Mach numbers. With an increasing temperature gradient the effective Mach number in the lower half of our domain has a steeper drop such that the stabilising effect of compressibility vanishes. Therefore, the stabilising effect of greater Mach numbers in a strongly stratified atmosphere is weaker.   
\subsection{Small P\'eclet number regime}
\label{subsec:Peclet_number_var}
\begin{figure}
\centering
   \includegraphics[width=\hsize]{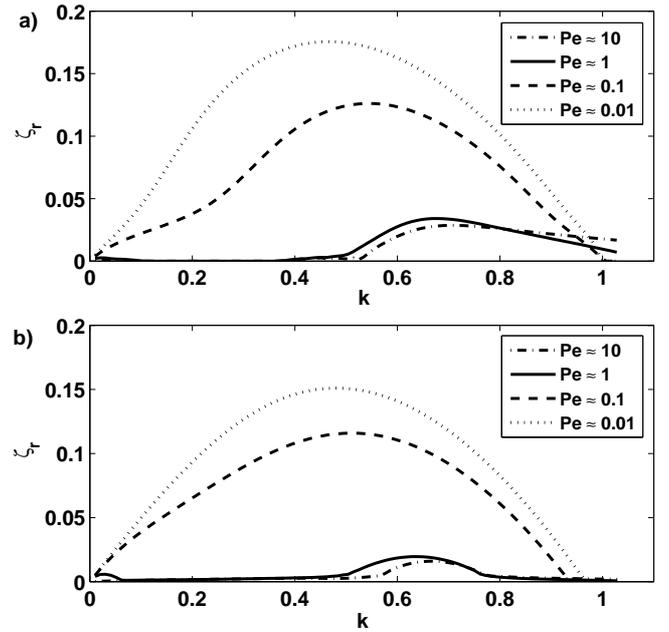}     
   \caption{Both plots show the growth rates $\zeta_{ r}$ in an inviscid weakly thermally stratified atmosphere with $\theta = 0.5$ for a Richardson number of $Ri= 0.22$. In a) the fluid is compressible with $M=0.09$ and in b) $M=0.009$, which corresponds to a incompressible fluid. The P\'eclet number is varied among three orders of magnitude.   
              }
         \label{fig:inviscid_different_ck}
   \end{figure}

\begin{figure}
\centering
   \includegraphics[width=\hsize]{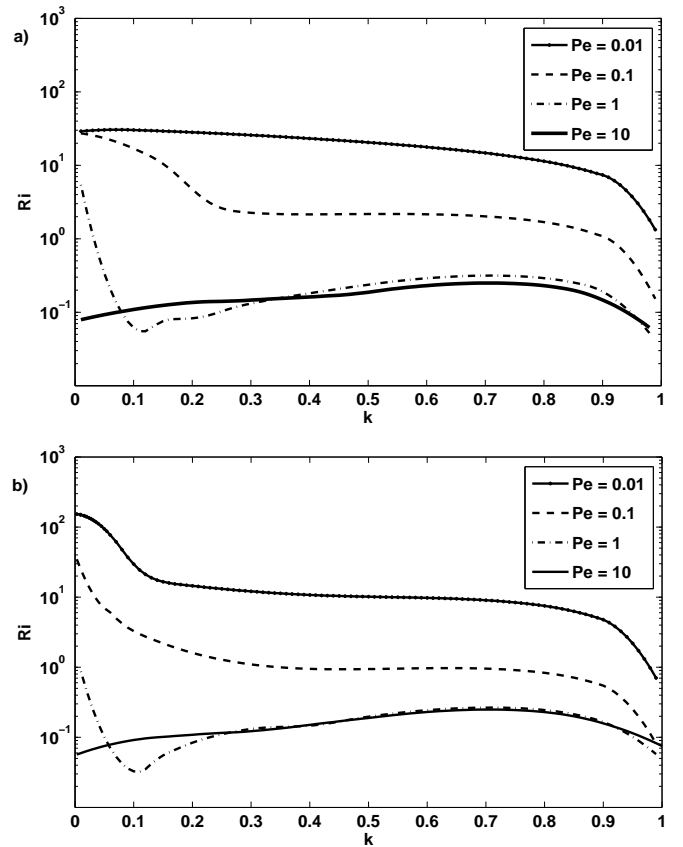}     
   \caption{Plots of critical Richardson numbers for all k in a viscous fluid with a Mach number, $M = 0.009$, a) $\theta = 0.5$ and b) $\theta = 2.0$ and viscosity ($\sigma C_k$) of order $10^{-7}$. The P\'eclet number is varied from 10 to 0.01.}
         \label{fig:viscid_different_ck_Ric_01}
   \end{figure}

In the following we focus on the impact of thermal diffusion in a stably stratified fluid. For non-negligible thermal diffusion the non-dimensional P\'eclet number is given as 
\begin{equation}
Pe = \frac{U_0 L_u}{C_k} \, ,
\end{equation}
where $L_u$ is the characteristic length of the shear width and $U_0$ correspond to the Mach number at the top of the domain as the velocity is normalised with respect to the sound speed. The P\'eclet number is associated with the ratio of advective transport to thermal diffusion.\\
Varying the P\'eclet number in an inviscid compressible fluid enhances the results found by \citet{1999AA...349.1027L} where a higher thermal diffusion destabilises the system as it effectively weakens the stable stratification, i.e. the system becomes more unstable against buoyancy. Therefore, thermal diffusion becomes important in a system where $t_{k} < t_{B}$, such that buoyancy is much slower than thermal diffusion. As $t_{B}$ has to be smaller than the system's dynamic time scale $t_S$, the P\'eclet number has to become smaller than unity to satisfy these requirements.\\
In Fig. \ref{fig:inviscid_different_ck} the growth rates for different P\'eclet numbers in a compressible and a weakly compressible fluid are illustrated. Comparing instability growth rates for a set up with $Pe < 1$ and $Pe > 1$ shows that $\zeta_{r}$ increases with decreasing $Pe$, where a significant jump can be observed when $Pe$ becomes smaller than unity. 
We find that the overall growth rate is smaller for a shear flow with Mach number equal to $0.09$, this demonstrates a stabilising effect of moderate Mach numbers as discussed in Sect. (\ref{subsec:low_temperature_gradient}). However, an increase in the instability growth rate does not necessary indicates a shift of the stability threshold to greater Richardson numbers.\\[10pt]
Focusing on the stability threshold it is necessary to seek for the critical Richardson numbers in different P\'eclet number regimes. For previous calculations of the growth rates, the fluid was assumed to be inviscid to simplify the problem and to avoid the issue of the initial state being actually not an equilibrium state. Here, it is more convenient to include viscosity as it is numerically easier to obtain results for the limit of small viscosity than for an inviscid fluid.\\ 
For a shear flow the onset of instability does not change for small enough $k$, if large enough Reynolds numbers are considered. However, including viscosity may have non-trivial effects on the stability if the fluid has large thermal diffusivity \cite[see][]{Jones1977}. \citet{Jones1977} derived a criterion for instability at long wavelength of the form $k Pe Ri < 0.086$ in an inviscid fluid. For a viscous fluid it can be rewritten as $k \sigma Re  Ri < 0.086$ such that for reasonably small wave numbers, the system can be still unstable if the Reynolds number is not sufficiently large. It can be partly seen when investigating equation (\ref{eq:EVProblemsystem02}) and (\ref{eq:EVProblemsystem03}), where almost all terms include the wave number, while one of the three viscous terms does not. Therefore, for small wave numbers this viscous term, which is proportional to the second derivative in $z$ direction, becomes relatively more important for the dynamics of the system.
To make sure that the results obtained correspond to a regime where viscosity does not affect the stability threshold, several computations were repeated with a smaller viscosity and a greater viscosity than used in the actual computations.\\
In Fig. \ref{fig:viscid_different_ck_Ric_01} curves of marginal stability for four different P\'eclet numbers in a polytropic atmosphere are displayed for $\theta = 0.5$ and $\theta = 2.0$. As expected the domain for an unstable shear flow increases as the P\'eclet number is decreased. It shows that not only is the growth rate of the instability altered but the stability threshold also changes for small P\'eclet numbers. The critical Richardson numbers for $Pe \leq 1 $ reveal a destabilisation for small $k$.  \citet{1999AA...349.1027L} explained this behaviour by the effect of the anisotropy of the buoyancy force. The stabilising effect of stratification becomes inefficient for predominant horizontal motion compared to the thermal diffusion. Indeed, by computing the ratio of the vertical to horizontal kinetic Energy of the unstable mode for a certain $k$ i.e.
\begin{equation}
\frac{E_{w}}{E_{u}}= \frac{\int_0^1 w_k(z)^2 dz}{\int_0^1 u_k(z)^2 dz},
\end{equation}
where $w_k(z)$ and $u_k(z)$ are the eigenfunctions for the vertical and horizontal velocity disturbances at a certain $k$ respectively, we are able to investigate the nature of the instability. Comparing this ratio for a mode with $k=0.1$ and $k=0.7$ for $Pe = 0.1$, it turns out that the ratio for the larger $k$-mode is of two orders of magnitude greater than for the smaller $k$-mode. Thus the horizontal motion associated with larger wave lengths in horizontal direction is predominant at very small $k$.\\
The results carried out for a greater temperature gradient, $\theta = 2$ shown in Fig. \ref{fig:viscid_different_ck_Ric_01} b, reveal a qualitatively different behaviour for the small P\'eclet numbers than it is the case for $\theta = 0.5$ shown in Fig. \ref{fig:viscid_different_ck_Ric_01} a. The $\theta = 0.5$ case has an overall greater critical Richardson number, but has a gradual increase towards a smaller critical Richardson number in the small $k$ limit. This indicates a more efficient destabilisation for most of the wave numbers and a less efficient destabilisation for small $k$. \\
Looking again at the ratio of kinetic energy in vertical and horizontal motions reveals that for lower $\theta$ the ratio of vertical motions to horizontal motions remains significantly smaller than for the higher stratification $\theta = 2$, which indicates that buoyancy force is more efficient in a strongly stratified atmosphere against the destabilising effect of thermal diffusion. A weakly stratified atmosphere can be destabilised quicker by thermal diffusion.
%
\subsubsection{Effect of the distance to the onset of convection on the instability}
\label{sec:varying_m}
\begin{figure*}
\centering
\includegraphics[width=\hsize]{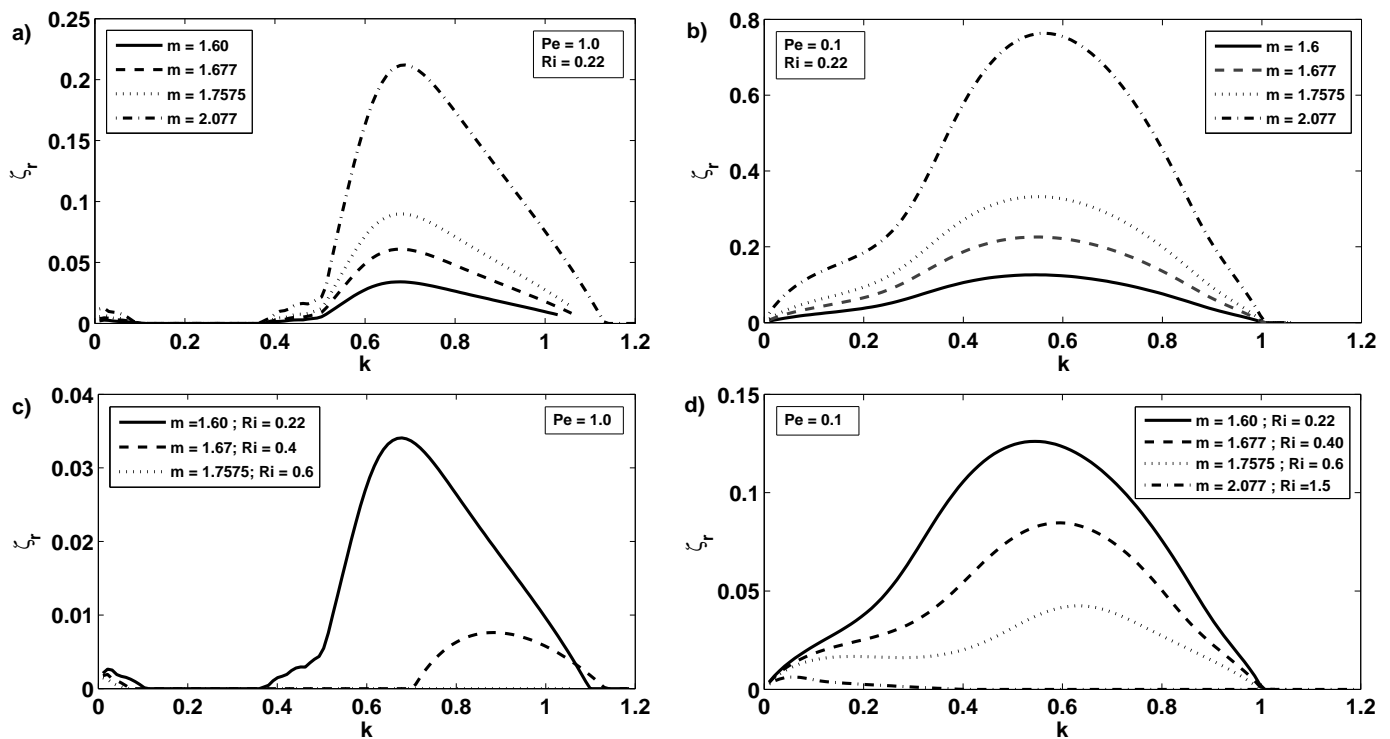}   
    \caption{For an inviscid fluid with thermal stratification $\theta = 2.0$ and a flow of low Mach number of order $10^{-2}$, the growth rate of the linear instability is plotted for several parameters. In  a) and c) the P\'eclet number is equal to unity and  in  b) and d) $Pe = 0.1$. $Ri$ is varied at the two bottom plots c) and d) but it is fixed to $Ri = 0.22$ for the top plots a) and b).}
         \label{diff_dist_cm}
   \end{figure*}

The destabilising mechanism of thermal diffusion can be traced back to the fact that thermal diffusion weakens the stable stratification against buoyancy. To investigate if, and how, this effect changes if the system is further away from the onset of convection the polytropic index is varied between 1.6, which is close to the threshold, and 2.07 which is far from the onset of convection.
In Fig. \ref{diff_dist_cm} a) the growth rate, $\zeta_{r}$, is plotted for different polytropic indices while the Richardson number is fixed to the value $Ri= 0.22$ and the P\'eclet number is unity. The growth rate is given in units of the speed sound, such that a rescaling is necessary while the amplitude of the shear flow is adjusted to keep $Ri$ constant. The same is done for a P\'eclet number of $0.1$ in Fig. \ref{diff_dist_cm} b) where the same tendency of an increasing growth rate with increasing polytropic index is observed. To exclude that this behaviour is due to the effect of thermal diffusion the growth rate for two different $m$ is computed for a fixed Ri smaller $1/4$ for much greater $Pe$ than unity, where the same tendency is found. However, the critical $Ri$ increases as $m$ is varied upwards.
Therefore, a system with larger polytropic index, $m$, is less stable such that for the same $Ri$ the system with a larger $m$ is further away from the stability threshold and has a higher growth rate. Non-linear direct numerical computations of two cases approved the results obtained with the linear EV solver. For this purpose equations (\ref{eq:NSEquation01}) - (\ref{eq:NSEquations}) were solved by using a hybrid finite-difference/pseudo-spectral code \cite[see for example][and references therein]{FLM:340261, 2009MNRAS.400..337S, 2009ApJ...702L..14S}. \\
While increasing the polytropic index requires a greater shear flow amplitude to obtain the same ratio between the buoyancy force and the turnover rate, the available kinetic energy of the system is increased. Consequently, as soon as the system can overcome the stabilising effect of density stratification the instability has more available kinetic energy from which it can grow more rapidly.\\
In the two plots at the bottom of Fig. \ref{diff_dist_cm} the growth rates for different polytropic indices are shown while all other parameter are held. The solid lines in Fig. \ref{diff_dist_cm} c) and d)  correspond to the solid lines in Fig.  \ref{diff_dist_cm} a) and b) respectively. As expected the instability growth rate decreases with increasing density stratification and greater $Ri$. 
While for a P\'eclet number of unity the instability shuts down rapidly, which is displayed in Fig. \ref{diff_dist_cm} c), for $Pe < 1$ the same behaviour is observed, but the instability for greater $m$ is still present for small wave numbers. As discussed in Sect. \ref{subsec:Peclet_number_var} this is explained by the anisotropy of the buoyancy force.
\subsection{Subdominant shear instability}
\label{subsec:Holmboe}
\begin{figure}
\centering
   \includegraphics[width=\hsize]{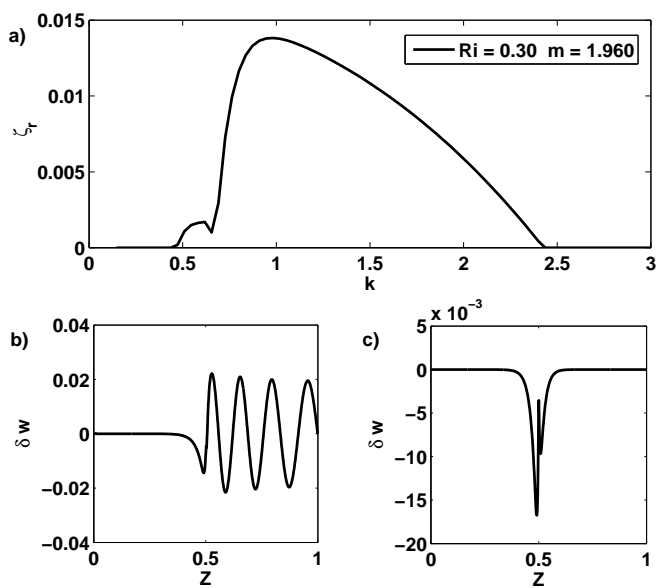}   
   \caption{The growth rate for an inviscid fluid with a temperature gradient, $\theta = 1.0$, P\'eclet number of order $10^3$ and $Ri= 0.3$ is shown in plot a). The form of the eigenfunctions found for the vertical velocity $w$ in the small arch and the large arch are displayed in b) and c) respectively.}
         \label{fig:Holmboe_instability}
\end{figure}

For certain configurations where the shear width is sufficiently small the velocity profile is similar to two counter flows. Then, for a stable stratification and P\'eclet numbers much greater than unity a Holmboe like instability is found. The Holmboe instability,  which was introduced by \citet{Holmboe_1962}, results from interacting waves, that propagate in opposite directions \citep{baines1994mechanism}. 
The Holmboe instability differs from the KH instability in several ways \cite[see the review article by][]{doi:10.1146/annurev.fluid.35.101101.161144}. First, the KH instability is stationary in a frame of reference, but the Holmboe instability has counter propagating unstable modes, which both have the same growth rate. One mode occupies the upper plane and the other the lower plane, such that a superposition of both solutions form a standing wave solution. Second, the Holmboe instability is favoured in stable stratified atmospheres and can dominate the Kelvin-Helmholtz instability when the stratification stability is increased, such that a KH instability disappear due to the Richardson criterion \citep{Smyth_peltier_holmboe}.\\  
In Fig. \ref{fig:Holmboe_instability} a) there are two arches present. The one with the greater maximum growth rate appears at greater $k$ and corresponds to the KH instability, where only one unstable mode is present for a certain wave number and its phase velocity corresponds to the mean flow velocity. The eigenfunction for the vertical velocity of these modes is shown in Fig. \ref{fig:Holmboe_instability} c), which reveals that the instability is localised at the center of the shear flow. Unstable modes with $k$ corresponding to the smaller arch exhibit all properties of the Holmboe instability, where two counter propagating modes, one in the upper and one in the lower half plane, are present. Fig. \ref{fig:Holmboe_instability} b) shows the form of the vertical velocity eigenfunction that is propagating in the lower half plane, the corresponding mode with the opposite phase velocity shows the oscillations in the upper half plane. Investigating all possible eigenfunctions at the overlap region between the two arches, reveals that Holmboe modes are present but have a smaller growth rate as the KH instability or \textit{vice versa}.\\
The observed Holmboe like instability only appears for large P\'eclet numbers while for small $Pe$ the range for the KH instability is enlarged to smaller wave numbers, such that even in the presents of the Holmboe like instability the KH instability for the same wave numbers dominates. This can be seen in Fig. \ref{diff_dist_cm} a) and b), where a small shoulder is present in the range $0.35 < k < 0.5$ in a) for $Pe = 1$ is dominated by the KH instability in b) for $Pe = 0.1$.\\
To make sure that these modes are physical and not a numerical artifact we checked that they remain for a system where we assume an inviscid non diffusive incompressible fluid. In a second check, we consider a non-uniform grid distribution in the $z$ direction, with more resolution where the largest gradients are observed. With both approaches a second instability, with the same properties, was found at slightly smaller wave numbers, $k$, than the KH instability. Using non-linear computations both instabilities are found during the linear growth regime with similar growth rates and eigenfunctions, as predicted from the linear computations.
\section{Conclusions}
\label{C_Conclusion}
Shear flow instabilities, that may lead to turbulent motions, are important for the understanding of dynamics in stellar interiors and in other astrophysical flows.
Using linear stability analysis the onset of shear flow instability in a compressible polytropic atmosphere was investigated, where a detailed analysis of the effect of moderate Mach numbers, small P\'eclet numbers and different polytropic indices was carried out separately. \\
For a flow of moderate Mach numbers we found a stabilising effect that results in a critical Richardson number which is less than $1/4$, such that for a certain range of $Ri$ less than $1/4$ the shear flow remains stable to KH instability. 
This effect becomes significant for Mach numbers greater than $0.1$. However, the stabilising is weaker for strongly stratified atmospheres. As this effect is most relevant at moderate to high Mach numbers it can be neglected for most stellar regions, in particular for the tachocline. However, this result might be important in other astrophysical objects where high Mach numbers are common.\\
For fluids with high thermal diffusivity, where the P\'eclet number drops below unity, a destabilisation of the system can be shown. Our results are in agreement with \citet{1999AA...349.1027L}, where the regime of small P\'eclet numbers were examined in a Boussinesq fluid. We find a significantly greater growth rate for highly diffusive fluids as well as greater critical Richardson numbers for $Pe < 1$ compared to systems with large P\'eclet numbers. Generally, steeper temperature gradients lead to an overall stabilisation for small P\'eclet numbers. However, it should be noted that, for very small wave numbers we found that the opposite is true due to anisotropy of the buoyancy force.\\
An interesting result is that we find the possibility of Holmboe like instabilities present in polytropic atmopheres. This instability is dominated by the KH instability for small P\'eclet numbers, but is clearly present for large P\'eclet numbers.
While no setup was found where the Holmboe like instability dominates the KH instability it may be found in future investigations.\\
Studying more complex systems, which include key properties of stars, by means of a linear stability analysis provides a powerful tool to investigate their linear behaviour. However, the vertical length scale for which these shear flows can become unstable is far below the resolution of helioseismological techniques and as the perturbation's amplitude can be finite in the tachocline \cite[see][]{1998ThCFD..11..183M} the assumption of infinitesimal perturbations made in the stability analysis might lead to different results. To clarify the second issue it is necessary to investigate the stability properties of such flows by means of direct numerical computations, where the initial state is perturbed by perturbations with a finite amplitude. In addition to finite amplitude effects, the non-linear behaviour of unstable shear flows after a saturation is crucially important for the understanding of mixing in stellar interiors regardless of the nature of the initial instability.\\
Having established an understanding of the hydrodynamical problem, further non-linear investigations are underway to obtain a full picture.

\begin{acknowledgements}
      This research has received funding from STFC and from the School of Mathematics, Computer Science and Engineering at City University London. We would also like to thank the anonymous referee for the helpful comments.
\end{acknowledgements}

\bibliographystyle{aa}
\bibliography{bib01}


\end{document}